\documentclass{nature}
\usepackage[english]{babel}
\usepackage{blindtext}
\usepackage{epsfig}
\usepackage{subcaption} 
\usepackage{ae, aecompl}
\usepackage{helvet}
\usepackage{epsfig}
\usepackage{dcolumn}
\usepackage{color}
\usepackage[LGR, T1]{fontenc}
\usepackage[latin9]{inputenc}
\usepackage{amssymb}
\usepackage{amsmath}
\usepackage{mwe, tikz}
\usepackage{esint}
\usepackage{longtable}
\usepackage{url}
\usepackage{etoolbox}
\usepackage{multirow}
\newcommand\aap{Astron. Astrophys.,}
\newcommand\apj{Astrophys. J.,}
\newcommand\apjl{Astrophys. J. Lett.,}
\newcommand\apjs{Astrophys. J., Suppl.}
\newcommand\mnras{Mon. Not. R. Astron. Soc.,}
\newcommand\nat{Nature,}

\usepackage{xpatch}

\title{Synchrotron radiation in $\gamma$-ray burst prompt emission} 

\author{
Bing Zhang
}

\begin{document}

\maketitle

\begin{affiliations}
\item Department of Physics and Astronomy, University of Nevada, Las Vegas, NV 89154, USA. Email: zhang@physics.unlv.edu
\item Center for Gravitational Physics, Yukawa Institute for Theoretical Physics, Kyoto University, Kyoto, 606-8502, Japan.
\end{affiliations}

%\begin{linenumbers}

%% Abstract
%%
\begin{abstract}
Growing evidence suggests that synchrotron radiation plays a significant role in shaping the spectra of most $\gamma$-ray bursts. The relativistic jets producing them likely carry a significant fraction of energy in the form of a Poynting flux.
\end{abstract}

The electromagnetic spectra of $\gamma$-ray bursts (GRBs) are usually characterized by a mathematical function invoking an exponentially-connected two-power-law function first proposed in a paper by the Burst And Transient Source Experiment (BATSE) team led by David Band (1957-2009)\cite{band93}. In the special session in memory of Band at the 2009 Fermi Symposium, Josh Grindlay, his PhD advisor from Harvard University, challenged GRB theorists to ``find the physical meaning'' of the ``Band function'' in 10 years. 

In the GRB field there was never a lack of models to interpret data. The challenge is rather to ``identify'' than to ``find'' the physical meaning of the Band function. Indeed, two leading models proposed long ago could both roughly account for the the general shape of the spectra but both models have one critical flaw regarding  the low-energy  photon index of the Band function, denoted as $\alpha$. The measured values of $\alpha$ peak around $\sim -1$ with a broad distribution\cite{preece00,nava11}. The first model invokes synchrotron radiation\cite{meszaros94,tavani96} of the electrons accelerated in the energy dissipation regions (internal shocks or magnetic reconnection sites) to account for the observed $\gamma$-rays. However, for a typical GRB environment the electrons are in the so-called ``fast-cooling'' regime that demands\cite{sari98,ghisellini00} $\alpha= -1.5$.  This is too ``soft'' to account for the data. The second model invokes quasi-thermal emission from a relativistic fireball\cite{meszarosrees00,rees05}.  The model typically predicts\cite{beloborodov10,deng14b} $\alpha \sim +1.5$.  This becomes too ``hard''.

The debate between the two models continued for many years, which may be coined as a ``battle of $\alpha$''. The rise of the photosphere model was motivated by the inability of the synchrotron model to account for $\alpha > -2/3$, the so-called ``synchrotron line-of-death''\cite{preece98}, as observed in a small fraction of bursts\cite{preece00,nava11}. The inclusion of a quasi-thermal component can significantly harden the GRB spectra\cite{meszarosrees00} to account for these bursts. The detection of a bright, thermally-dominated burst, GRB 090902B\cite{abdo09b,ryde10}, offered confidence to the photosphere modelers, who speculated that by spatially and temporally superposing photosphere emission from different parts of the jet, much softer $\alpha$ can be also obtained within the photosphere models. It was until detailed modeling when people realized that the superposition effect has limited effect on softening $\alpha$\cite{deng14b,parsotan18}. Contrived structured jet configurations (e.g. structured Lorentz factor profile but uniform luminosity profile) are needed in order to soften $\alpha$ significant enough to meet the data requirements\cite{lundman13,meng19}.  

On the other hand, synchrotron modelers over the years have proposed various scenarios to harden the spectrum from the fast-cooling $\alpha = -1.5$ to the typical $\alpha \sim -1$\cite{peerzhang06,daigne11,uhm14,xu17}. However, it takes very contrived conditions to break the $\alpha=-2/3$ line-of-death limit\cite{yangzhang18}. One step forward was the realization that one should use the synchrotron model spectrum itself (rather than the Band function) to fit the data directly\cite{burgess14,burgess15,zhangbb16a}. Somewhat surprisingly, seemly different models (e.g. synchrotron vs. Band) could lead to similar goodness of fitting to the same set of data\cite{zhangbb16a}. This is because the limited number of $\gamma$-ray photons  detected from a GRB allows the flexibility to match different input models. Applying physical synchrotron models to fit the GRB data turned out to be fruitful: Besides proving that modified fast-cooling models can fit the data\cite{zhangbb16a}, the standard fast cooling spectrum may also interpret some GRBs given that there exists an additional break in the even lower energy regime\cite{zheng12,oganesyan17}. This second break is indeed found to exist in a good fraction of GRBs\cite{ravasio19,oganesyan19}. More intriguingly, the synchrotron model can also account for the majority of the bursts that are beyond the traditional line-of-death\cite{burgess19}, suggesting that the $\alpha = -2/3$ line for the Band function model no longer carries a significant physical meaning. According to Ref. \cite{burgess19}, essentially all the GRB spectra can be accounted for within the synchrotron model except very few cases.

This does not mean that the photosphere emission is not shaping the observed GRB spectra. The spectra of some bursts are best fit with the superposition of a Band component and a thermal component\cite{guiriec11,axelsson12,guiriec13}, suggesting the co-existence of the photosphere component and the synchrotron component. The spectra of a handful of  GRBs (e.g. GRB 090902B\cite{abdo09b,ryde10}) are dominated by the quasi-thermal photospheric emission, but these are exceptions rather than norms.

The diverse GRB spectra can be understood within the framework of a diversity of jet composition among GRBs. The ultimate energy power source of a GRB comes from two energy reservoirs (Figure 1)\cite{zhang18}: The first is the gravitational energy of the engine which is released as the huge thermal energy in a ``fireball''; the second is the rotational energy of the engine, which is tapped via a Poynting flux. A fraction of the thermal energy escapes from the photosphere, powering the thermal component in the GRB spectra. Both the thermal energy and the Poynting flux energy can be converted into the kinetic energy of the outflow, which can be dissipated in internal shocks, leading to particle acceleration and synchrotron radiation. The Poynting flux energy can be also directly dissipated through magnetic reconnection, also leading to particle acceleration and synchrotron radiation.

Whether the spectrum of a GRB is synchrotron-dominated or thermally-dominated depends on the initial allocation of energy in the two energy reservoirs at the central engine\cite{gaozhang15}. The standard fireball usually predicts a more significant thermal component\cite{meszarosrees00}, as observed in GRB 090902B\cite{abdo09b,ryde10}. The non-detection of such a bright thermal component in the majority of the bursts suggests that most bursts have a significant Poynting flux component. The GRB jet composition is likely hybrid, with the ``fireball'' case  and the pure ``Poynting-flux-dominated'' case as the two extreme regimes (Fig.1).  The majority of GRBs have both energy components, with the Poynting flux component dominating in the central engine and sometimes in the emission region as well. 

A consensus of the synchrotron modelers is that in order to reproduce the data, the emission region needs to be at least two orders of magnitude farther away from the engine than the standard internal shock site\cite{kumarmcmahon08,beniamini13,uhm14,oganesyan17,ravasio19,burgess19}. Such large radii for synchrotron radiation are expected in the Poynting-flux dissipation models invoking repeated collisions\cite{zhangyan11} or collision- or deceleration-induced magnetic kink instabilities\cite{lazarian19}. The large emission radius of the GRB prompt emission is also consistent with the non-detection of high-energy neutrinos from any GRBs\cite{icecube17,zhangkumar13} and the modeling of the ``spectral lags'' and spectral evolution patterns in GRB pulses\cite{uhm16b,uhm18}. The dissipation of a Poynting flux in GRBs is evidenced by the detection of moderate linear polarization in both prompt $\gamma$-ray\cite{yonetoku11,zhangsn19}  and optical\cite{troja17b} emissions as well as the early optical emission  believed to originate from the reverse external shock\cite{steele09,mundell13}.

So, after a decade from 2009, the main component that shapes the GRB prompt emission spectra is identified\cite{burgess14,zhangbb16a,oganesyan17,ravasio19,burgess19}. With detailed observations and data analyses of the broad-band prompt emission data ($\gamma$-rays, X-rays and optical) of many GRBs using the current and future GRB observatories, it would be possible to quantify the distribution of the GRB jet composition in the future. After all, the best tribute to David Band would be to replace the Band function by more physical models to understand the ever-growing GRB data in the years to come.

%\bibliography{ms}

%\end{linenumbers}

\newpage

\begin{figure}[t]
\begin{tabular}{c}
\includegraphics[keepaspectratio, clip, width=0.6\textwidth]{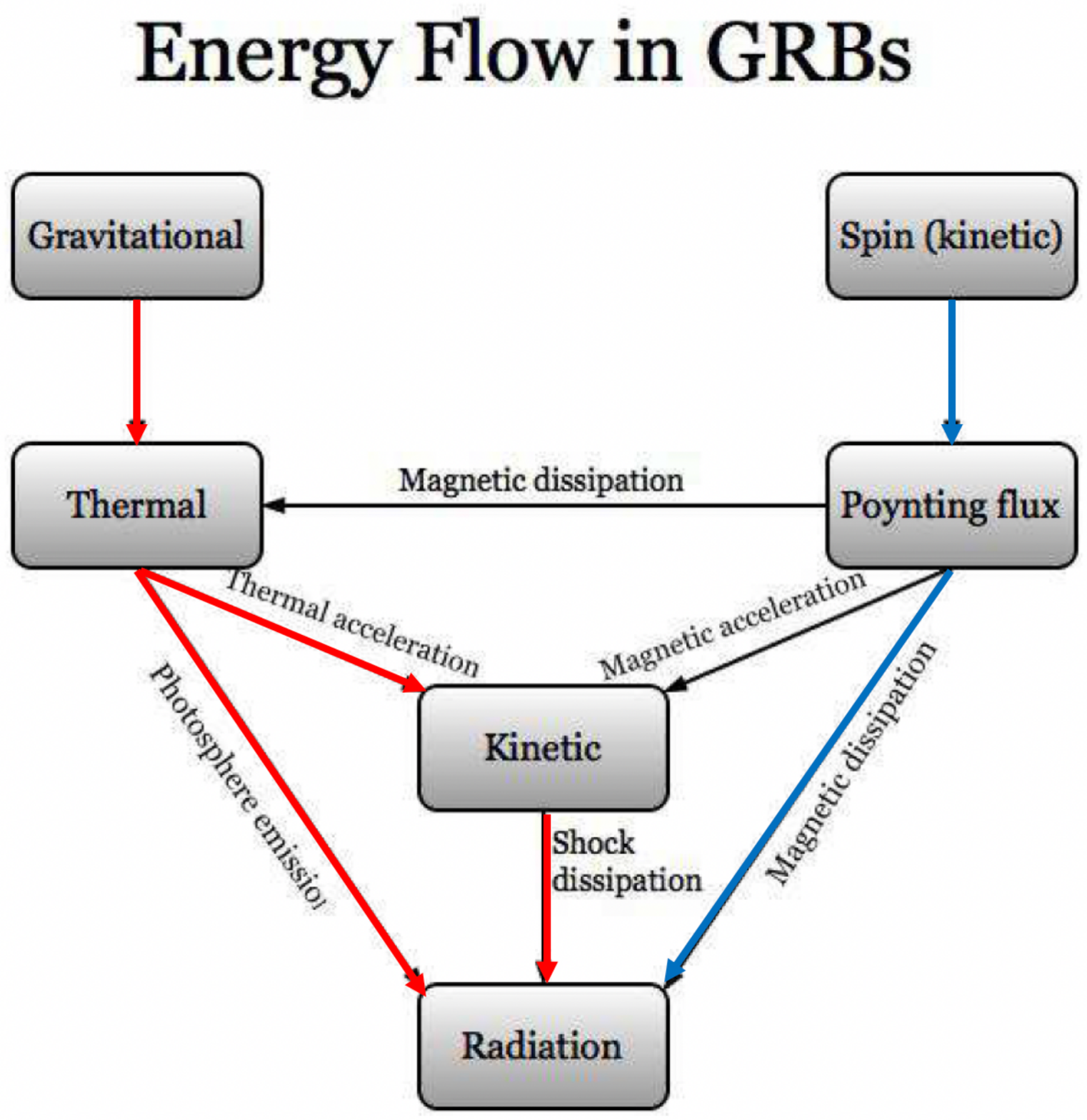} \\
\end{tabular}
\caption{{\bf The energy flow chart in GRBs\cite{zhang18} for different prompt emission models}. The standard fireball model (red arrows)\cite{meszarosrees00} invokes a thermally-dominated energy at the central engine, which is converted to the kinetic energy and subsequently dissipated in internal shocks. The model predicts a bright photosphere emission component and a synchrotron component from the internal shocks. The Poynting-flux-dominated model (blue arrows) can directly dissipate the Poynting flux energy to power bright synchrotron radiation\cite{zhangyan11}. A realistic GRB likely invokes both energy reservoirs at the central engine, so that the jet composition is likely hybrid\cite{gaozhang15}. All the arrows in the flow chart are relevant. The dominance of each emission component depends on the initial allocation of the energy budget from the two energy reservoirs.}
\label{fig:lc_sp}
\end{figure}

\end{document}